\documentclass[aps,prd,twocolumn,nofootinbib,showpacs]{revtex4}

\usepackage{graphicx}
\usepackage{dcolumn}
\usepackage{bm}
\usepackage{epsfig}

\begin{document}

\title{Timelike Form Factor of the Kaon for $|Q^2|=M^2(J/\psi)$}

\author{Kamal K. Seth}

\affiliation{Northwestern University, Evanston, Illinois 60208}


\begin{abstract} 
The timelike form factor of the charged kaon is evaluated at $|Q^2|=M^2(J/\psi)=9.59~\mathrm{GeV}^2$ by relating it to $\mathcal{B}(J/\psi\to K^+K^-)$, $\mathcal{B}(J/\psi\to K_SK_L)$ and $\mathcal{B}(J/\psi\to e^+e^-)$.  The resulting value, $|Q^2|F_K(9.59~\mathrm{GeV}^2)=0.81\pm0.18~\mathrm{GeV}^2$ is found to be in good agreement with  $|Q^2|F_K(13.5~\mathrm{GeV}^2)=0.85\pm0.05~\mathrm{GeV}^2$ obtained recently by the direct measurement of the reaction $e^+e^-\to K^+K^-$.
\end{abstract}

\maketitle

Electromagnetic form factors provide direct insight into the electromagnetic structure of a hadron, the distribution of charges and currents in the hadron as they couple with the photon.  Since targets of unstable hadrons are not possible, determination of their from factors for spacelike momentum transfers (positive $Q^2$) at large momentum transfers becomes all but impossible.  However, form factors of a hadron ($h$) for timelike momentum transfers (negative $Q^2$) can be measured by $e^+e^-\to h^+h^-$ reactions.  Unfortunately, until recently such measurements for pions or kaons were rare, and for $|Q^2|>4.7~\mathrm{GeV}^2$ they had either upper limits or results with errors $\ge50\%$ \cite{literature}.  Recently, CLEO \cite{pete} reported measurements of $|Q^2|F_\pi$ and $|Q^2|F_K$ for $|Q^2|=13.48~\mathrm{GeV}^2$ with errors of $\sim\pm15\%$ and $\pm5\%$, respectively, providing the first opportunity to critically test theoretical predictions.  Since these measurements for a single value of $|Q^2|$ are far removed from accurate measurements at smaller momentum transfers, they do not shed light on the variation of $|Q^2|F_{\pi,K}$ with $|Q^2|$, which is a distinguishing characteristic of theoretical predictions.  Sometime ago, Milana, Nussinov and Olsson \cite{milana} provided an \textit{estimate} of  $|Q^2|F_\pi(9.59~\mathrm{GeV}^2)$ by relating the measurements of $\mathcal{B}(J/\psi\to\pi^+\pi^-)$ and $\mathcal{B}(J/\psi\to e^+e^-)$ to the form factor at $M^2(J/\psi)=9.59~\mathrm{GeV}^2$.  The result of Milana~et~al.~\cite{milana}, modified for the PDG06 \cite{pdg} value of $\mathcal{B}(J/\psi\to e^+e^-)$, is $|Q^2|F_\pi(9.59~\mathrm{GeV}^2)=0.95\pm0.07~\mathrm{GeV}^2$.  The excellent agreement of this result with the result of the recent CLEO measurement \cite{pete} $|Q^2|F_\pi(13.48~\mathrm{GeV}^2)=1.01\pm0.13~\mathrm{GeV}^2$ provides remarkable confirmation of the validity of the arguments provided by Milana et al. In this communication we use the same line of arguments to obtain $|Q^2|F_K(9.59~\mathrm{GeV}^2)$, and thus obtain the only determination of the kaon form factor at a large momentum transfer, other than that from the CLEO measurement \cite{pete}.

\begin{figure*}[!tb]
\begin{center}
\includegraphics[width=5.in]{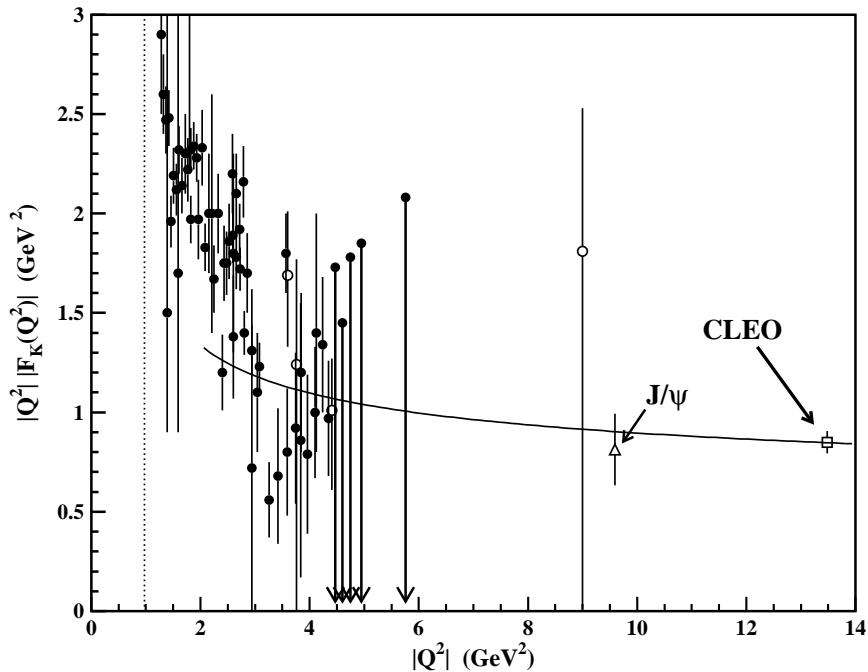}
\end{center}
\caption{Compilation of the measurements of $|Q^2|F_K(Q^2)$ based on Ref.~\cite{literature}, the CLEO measurements \cite{pete}, and the present result for $|Q^2|=M^2(J/\psi)$.  The solid curve shows the arbitrarily normalized variation of $\alpha_S(Q^2)$.}
\end{figure*}

As is well known, $J/\psi$ annihilation into a meson pair ($M\overline{M}$) can proceed by three different intermediaries: a photon, three gluons, or two gluons plus a photon.  In other words,
\begin{equation}
\mathcal{B}(J/\psi\to M\overline{M})=K|A_\gamma + A_{ggg} + A_{\gamma gg}|^2
\end{equation}
where $K$ is a constant.  Milana et al. showed that for the production of a pion pair $A_{ggg}$ and $A_{\gamma gg}$ are small and can be neglected.  The decay therefore proceeds essentially exclusively via a virtual photon, so that one can write
\begin{equation}
\frac{\mathcal{B}(J/\psi\to\pi^+\pi^-)}{\mathcal{B}(J/\psi\to e^+e^-)}=2F_\pi^2(M^2_{J/\psi})\times\left(\frac{p_\pi}{M_{J/\psi}}\right)^3
\end{equation}
where the pion phase space factor, $(p_\pi/M_{J/\psi})^3=0.125$.

In order to determine the kaon form factor at $|Q^2|=M_{J/\psi}^2$, we have to revisit the arguments of Milana~et~al. for the relative importance of the three amplitudes in Eq.~1.

The pQCD prediction \cite{pqcd} including the lowest order radiative correction is
\begin{equation}
\frac{\Gamma_{\gamma gg}}{\Gamma_{ggg}} = \left( \frac{16\alpha_{em}}{5\alpha_S}\right) \left( \frac{1-6.7\alpha_S/\pi}{1-3.7\alpha_S/\pi}\right) = 0.029,
\end{equation}
using $\alpha_S(J/\psi)=0.35$.  Milana et al. have argued that for the purposes of the form factor calculation in which the final state is pure hadrons, the virtual photon in the $\gamma gg$ intermediate state must convert into a $q\bar{q}$ pair, so that the above ratio is reduced by an additional factor $\alpha$.  Thus, $\Gamma_{\gamma gg}(\gamma\to q\bar{q})/\Gamma_{ggg} \approx 3\times10^{-4}$, and $A_{\gamma gg}$ in Eq. 1 can be neglected.  This argument completely carries over to the $K^+K^-$ decay of $J/\psi$, and we obtain
\begin{equation}
\mathcal{B}(J/\psi\to K^+K^-) = K |A_\gamma + A_{ggg}|^2.
\end{equation}
By independent analyses of the available experimental data, Suzuki \cite{suzuki} and Rosner \cite{rosner} have shown that for $J/\psi$ decay into two pseudoscalars, the two amplitudes $A_\gamma$ and $A_{ggg}$ are nearly $90^\circ$ out of phase.  This result is confirmed by Bai et al. in an analysis including the latest experimental data from BES~\cite{bes}. As a result, Eq.~4 leads to
\begin{eqnarray}
 \mathcal{B}_\gamma(J/\psi\to K^+K^-) \hspace{2.in}\\
\nonumber = \mathcal{B}(J/\psi\to K^+K^-) - \mathcal{B}_{ggg}(J/\psi\to K^+K^-)
\end{eqnarray}
As is well known \cite{rosner,suzuki}, charge conjugation does not allow the decay of a flavor SU(3) singlet state of $J^{PC}=1^{--}$ into a $0^-0^-$ pair.  The gluonic decay therefore may only proceed via the SU(3) breaking part of the pseudoscalars.  For the pions this part is negligibly small, but it can be substantial for the kaons.  The decay to $K_SK_L$, having no direct photon part, proceeds only through this SU(3) breaking part.  We therefore identify $\mathcal{B}_{ggg}(J/\psi\to K^+K^-)$ with $\mathcal{B}(J/\psi\to K_SK_L)$. Eq.~5 then becomes
\begin{eqnarray}
\mathcal{B}_\gamma(J/\psi\to K^+K^-) \hspace{2.in}\\
\nonumber = \mathcal{B}(J/\psi\to K^+K^-) - \mathcal{B}(J/\psi\to K_SK_L).
\end{eqnarray}
Thus, corresponding to Eq.~2, we get
\begin{eqnarray}
\frac{\mathcal{B}_\gamma(J/\psi\to K^+K^-)}{\mathcal{B}(J/\psi\to e^+e^-)} = 2F_K^2(M^2_{J/\psi})\times\left(\frac{p_K}{M_{J/\psi}}\right)^3
\end{eqnarray}

Inserting the PDG \cite{pdg} values of the branching ratios, $\mathcal{B}(J/\psi\to K^+K^-)=(2.37\pm0.31)\times10^{-4}$, $\mathcal{B}(J/\psi\to K_SK_L)=(1.46\pm0.26)\times10^{-4}$,  $\mathcal{B}(J/\psi\to e^+e^-)=(5.94\pm0.06)\times10^{-2}$, and the phase space factor $(p_K/M_{J/\psi})^3=0.106$, we get \cite{note}
\begin{equation}
|Q^2|F_K(9.59~\mathrm{GeV}^2)=0.81\pm0.18~\mathrm{GeV}^2
\end{equation}
This is to be compared with the CLEO result of direct measurement
\begin{equation}
|Q^2|F_K(13.48~\mathrm{GeV}^2)=0.85\pm0.05~\mathrm{GeV}^2
\end{equation}
Again, it is reassuring that the two results are very close.  This adds to our confidence in the arguments used in deriving the above result for $|Q^2|F_K(M^2_{J/\psi})$.

We note that with the present result, the ratio $F_K(9.59~\mathrm{GeV}^2)/F_\pi(9.59~\mathrm{GeV}^2)=0.85\pm0.20$ is also in good agreement with the CLEO result, $F_K(13.48~\mathrm{GeV}^2)/F_\pi(13.48~\mathrm{GeV}^2)=0.84\pm0.12$.

Fig.~1 shows the compilation of all existing measurements of $|Q^2|F_K(|Q^2|)$. The curve, normalized at the CLEO result, illustrates the pQCD prediction that $|Q^2|F_K(|Q^2|)$ should vary as $\alpha_S(|Q^2|)$.  No theoretical predictions for $|Q^2|F_K(|Q^2|)$ are available for comparison.

The author wishes to thank Dr.~Peter~Zweber for discussions.  This work was supported by the U.S. Department of Energy.


\begin{thebibliography}{99}
\bibitem{note} It has been noted that the recent BES result \cite{bes} for $\mathcal{B}(J/\psi\to K_SK_L)=(1.82\pm0.14)\times10^{-4}$ is nearly 2.5$\sigma$ larger than the old Mark 3 result \cite{markiii} $\mathcal{B}(J/\psi\to K_SK_L)=(1.18\pm0.22)\times10^{-4}$.  The BES result is based on a very detailed analysis of 58 million $J/\psi$ which resulted in $2155\pm45$ $K_SK_L$ events.  The Mark 3 result is based on the analysis of 2.7 million $J/\psi$ which resulted in 74 $K_SK_L$ events.  Using only the BES resulted would lead to $|Q^2|F_K(9.59~\mathrm{GeV}^2)=0.63\pm0.19~\mathrm{GeV}^2$, and using only the Mark 3 result would lead to $|Q^2|F_K(9.59~\mathrm{GeV}^2)=0.93\pm0.15~\mathrm{GeV}^2$.  We believe it is prudent to use the PDG average \cite{pdg} whose error reflects the discrepancy between the old and the new results for $\mathcal{B}(J/\psi\to K^+K^-)$.
\end{thebibliography}
\end{document}